# The formation heritage of Jupiter Family Comet 10P/Tempel 2 as revealed by infrared spectroscopy

L. Paganini[1,2], M.J. Mumma[1], B.P. Bonev[1,3], G.L. Villanueva[1,3],
M.A. DiSanti[1], J.V. Keane[4], and K.J. Meech[4]

[1]Goddard Center for Astrobiology, NASA GSFC, MS 690.3, Greenbelt, MD 20771, USA
[2] NASA Postdoctoral Fellow, `lucas.paganini@nasa.gov`
[3] Department of Physics, Catholic University of America, Washington, DC 20064, USA
[4]Institute for Astronomy, Univ. of Hawaii, Honolulu, HI 96822

**Abstract**

We present spectral and spatial information for major volatile species in Comet 10P/Tempel 2, based on high-dispersion infrared spectra acquired on UT 2010 July 26 (heliocentric distance $R_h$ = 1.44 AU) and September 18 ($R_h$ = 1.62 AU), following the comet's perihelion passage on UT 2010 July 04. The total production rate for water on July 26 was $(1.90 \pm 0.12) \times 10^{28}$ molecules s$^{-1}$, and abundances of six trace gases (relative to water) were: $CH_3OH$ (1.58% ± 0.23), $C_2H_6$ (0.39% ± 0.04), $NH_3$ (0.83% ± 0.20), and HCN (0.13% ± 0.02). A detailed analysis of intensities for water emission lines provided a rotational temperature of 35 ± 3 K. The mean OPR is consistent with nuclear spin populations in statistical equilibrium (OPR = 3.01 ± 0.18), and the (1σ) lower bound corresponds to a spin temperature > 38 K. Our measurements were contemporaneous with a jet-like feature observed at optical wavelengths. The spatial profiles of four primary volatiles display strong enhancements in the jet direction, which favors release from a localized vent on the nucleus. The measured IR continuum is much more sharply peaked and is consistent with a dominant contribution from the nucleus itself. The peak intensities for $H_2O$, $CH_3OH$, and $C_2H_6$ are offset by ~200 km in the jet direction, suggesting the possible existence of a distributed source, such as the release of icy grains that subsequently sublimed in the coma. On UT September 18, no obvious emission lines were present in our spectra, nevertheless we obtained a 3σ upper limit $Q(H_2O) < 2.86 \times 10^{27}$ molecules s$^{-1}$.

*Subject headings:* Comets, composition; Comets, origin; IR spectroscopy; Organic chemistry; Prebiotic chemistry

1. **Introduction**

Given the complex scenario and dynamics that occurred during Solar System formation, and later during the evolution to the current planetary system, questions persist not only on the time, place, and conditions under which comets formed, but also on the degree to which cometary ices were processed after cometesimals aggregated.

Rigorous dynamical models have expanded this scenario by addressing the origin of comets in the Kuiper Belt (KB) and Oort Cloud (OC) reservoirs. The *Nice* model suggests that comets were expelled from the inner-Neptunian region (about 5–30 AU from the proto-Sun) to their actual present day reservoirs as a result of interactions with the giant planets (Gomes et al. 2005; Tsiganis et al. 2005; Morbidelli et al. 2008). A later adaptation explored the possible capture of Oort Cloud comets from stars in the Sun's birth cluster (Levison et al. 2010). Such models imply profound differences in the place of origin of pre-cometary materials, in their physical-chemical signatures, and in the composition of primitive icy planetesimals. Once formed and ejected to quiescent regions (KB, OC), the native composition of cometary nuclei is believed to remain isolated from external catalysts. If so, the nature of the volatile fraction in comet nuclei carries fundamental clues to these processes.

The molecular composition and taxonomy of comets provide essential information that might help us trace the primordial conditions of the Solar System. The large diversity of molecular abundances measured in comets, however, presents major constraints that must be satisfied when classifying these objects and placing them in the context of suitable formation scenarios.

Comet 10P/Tempel 2 (hereafter 10P) was discovered by Ernst Wilhelm Liebrecht Tempel in 1873. Since then, 10P has been characterized by several studies, especially during the 1988 apparition (e.g., A'Hearn et al. 1989; Jewitt and Luu 1989; Sekanina 1991; Sykes et al. 1998; Biver et al. 2002; Knight et al. 2011, and references therein). Light curve studies of 10P suggest a spheroidal nucleus of 16.3 × 8.6 × 8.6 km$^3$ with a rotation period of 8.9 hr. Its orbit is inclined to the ecliptic by 12.0°, the Jupiter Tisserand parameter with respect to Jupiter is 2.963, and the period is 5.37 years. Together these classify 10P as an ecliptic comet of the Jupiter dynamical family.

In this study, we report gas production rates, chemical composition, and spatial properties of primary volatiles in Jupiter family comet (JFC) 10P/Tempel 2. We compare water production rates measured during its 2010 and 1988 apparitions, and compare the chemical composition revealed herein with that of other JFCs characterized at infrared (IR) wavelengths. Ultimately, we attempt to distinguish possible formation scenarios for 10P, using cosmogonic parameters inferred from our observational results.

## 2. Observations and data analysis

We conducted astronomical observations of comet 10P using NIRSPEC at the 10-meter W. M. Keck Observatory (Keck II) atop Mauna Kea, Hawaii, on UT 2010 Jul 26 and Sep 18. NIRSPEC is a cross-dispersed echelle grating spectrometer which features a 1024 × 1024 InSb array detector with pixels that sample 0.144″ (spectral) × 0.198″ (spatial) extent (McLean et al. 1998). The cross-dispersed capability allows NIRSPEC to sample a wide range of spectral orders with relatively high spectral resolving power ($\lambda/\Delta\lambda \sim 25{,}000$), enabling the simultaneous detection of ro-vibrational transitions of multiple gases and their spatial distributions along the slit. In 10P we quantified rotational temperatures for water and HCN, production rates for six gases, and possible cosmogonic parameters such as the abundance ratio (OPR) of nuclear spin species (ortho, para) of water (and the equivalent spin temperature, $T_{spin}$) and the mixing ratios of primary (parent) volatiles.

We used a 0.432″ × 24″ slit configuration for observations of comet 10P. This arrangement allowed simultaneous sampling of six spectral orders within the L band (2.8–4.1 $\mu$m), in each order encompassing about 40 cm$^{-1}$. In July 2010, instrument setting KL1 sampled emission lines of $H_2O$, OH*, $C_2H_6$, and $CH_3OH$ while setting KL2 sampled $H_2O$, OH*, HCN, $H_2CO$, $C_2H_2$, and $NH_3$. We used only the KL2 setting for the September observations (see Table 1). Weather conditions were good on both dates. Seeing was 0.9″ (0.7″) FWHM and the airmass was 1.15 (1.5) on 2010 July 26 (September 18). The comet was observed using a four-step sequence ABBA with an integration time of 60 sec/step, nodding the telescope by 12″ between the A and B positions (i.e., a single ABBA sequence required 4 min integration time for completion). Since the comet was a relatively small object, it was contained entirely within the slit length (24″, or ∼120 pixels). The operation ($A_1-B_1-B_2+A_2$) provided cancellation (to second order, in a Taylor series expansion about the mean air mass) of thermal background emission and of "sky" line emission from the Earth's atmosphere.

We followed our standard methodology for data reduction and analysis of the individual echelle orders (e.g., see Bonev 2005 for details). Initial data processing included the removal of high dark current pixels and cosmic-ray hits, flat-fielding, spatial and spectral rectification, and spatial registration. After combining the A and B beams from the difference frames, we extracted a spectrum from the sum of 9 spatial rows (1.78″) centered on the nucleus (this also contributes to removing residual background, see Bonev 2005 Appendix 2). The extracted spectrum contains emission lines of cometary gases and continuum radiation associated with cometary dust grains. We isolated the emission lines by subtracting a transmittance function synthesized for the terrestrial atmosphere (normalized to the cometary continuum) from the combined spectrum. We synthesized the spectral transmittance by using a multiple layer atmosphere and a radiative transfer model (LBLRTM) that accessed the HITRAN-2008 molecular database enhanced with our custom updated line parameters and fluorescence models (e.g., Villanueva et al. 2011a). Flux calibration was achieved through observations of the IR standard stars BS0015 and BS1040 with a 0.720″ × 24″ slit configuration.

On 2010 July 26, we detected several lines of $H_2O$ and OH* (prompt emission) in the combined KL1 and KL2 settings (see Fig. 1), along with the presence of minor volatiles. Figure 2 shows spectral lines of ethane ($C_2H_6$) and methanol ($CH_3OH$) using the KL1 setting alone. In addition, the KL2 setting provided detections of hydrogen cyanide (HCN) and ammonia ($NH_3$), as well as upper limits for formaldehyde ($H_2CO$) and acetylene ($C_2H_2$). Even though the UT 2010 September observations did not show clear spectral, these data allowed us to obtain a sensitive upper limit for water production in 10P.

## 3. Results

After applying fluorescence models to these spectra, we retrieved consistent rotational temperatures (Section 3.1) and production rates (Section 3.2). In particular, Table 2 gives a summary of the observed volatiles, as well as their total flux, fluorescence efficiencies (or g-factors), rotational temperatures, production rates, and relative abundance (relative to water). Section 3.3 presents a determination of the ortho-to-para ratio and spin temperature in water. Spatial profiles for $H_2O$, $C_2H_6$, $CH_3OH$, HCN and the continuum are discussed in Section 3.4.

### 3.1. Rotational temperature

Because the rotational populations are not completely sampled, an accurate measure of the rotational temperature ($T_{rot}$) is needed to estimate the complete population of a particular species from the lines measured, i.e., an accurate value for $T_{rot}$ is essential for retrieving accurate production rates in the coma. We obtain $T_{rot}$ by correlation and excitation analyses using precise quantum band models. For water, we used ortho-water lines extracted from KL1 and KL2 settings (Figs. 1 and 3). We compared the extracted line intensities with those predicted by an advanced $H_2O$ fluorescence model, synthesized at intervals of 1 K. The $H_2O$ model employs a combination of existing spectral databases to compute detailed (non-resonance) fluorescence of many (hot-band) water transitions (Villanueva et al. 2012).

The resulting $H_2O$ rotational temperature, $T_{rot}$ (35 ± 3 K), applies to our nucleus-centered aperture of 3 by 9 pixels (equivalent to $0.43''$ × $1.78''$, or 266 × 800 km on July 26). The value derived for HCN ($T_{rot} = 36^{+13}_{-9}$ K), based on ten emission lines (see Fig. 3 and Table 2), is consistent with the rotational temperature obtained for $H_2O$.

### 3.2. Production rates and molecular abundances

An apparent production rate is determined from the flux of each ro-vibrational transition detected within our sampled aperture, $F_{line}$ [W m$^{-2}$]. Our methodology considers the molecular lifetime, $\tau$ [s], and the fraction of total molecular content in the coma sampled by each pixel, $f(x)$. Assuming a constant outflow velocity (0.8 $R_h^{-0.5}$ km

s$^{-1}$) and uniform decay rate in the coma, the product of $\tau$ and *f(x)* is independent of molecular lifetime. Fluorescence efficiencies are based on quantum-mechanical models for each molecule (see Villanueva et al. 2011b, and reference therein). Current astrometric parameters are accounted for at the time of our observations, i.e., geocentric distance ($\Delta$), the terrestrial transmittance ($T_i$), and the proper fluorescence efficiency (the so-called "g-factor", $g_i$) at the appropriate rotational temperature and heliocentric distance, $R_h$. The nucleus-centered production rate derived from the individual line is given by:

$$Q_{nc} = \frac{4\pi\Delta^2 F_{line}}{\tau g_i T_i f_{(x)}} \quad .$$

In the process of obtaining production rates, a matrix of individual 0.144″ by 0.198″ extracts is used to calculate the net flux contribution within the sampled aperture (3 × 9 pixels). Individual line fluxes ($F_{line}$) are affected by "seeing" effects (blurring and twinkling due to the Earth's atmosphere) and by slight drifts of the comet during individual exposures of the ABBA sequences (see DiSanti et al. 2006, and references therein). So the nucleus-centered flux, smeared out by these phenomena, underestimates the real production rates observed. This issue, however, is compensated by a correction or *growth* factor (see Fig. 4), which essentially describes the ratio of terminal to nucleus-centered production rates (where *terminal* refers to off-nucleus extracts along the slit spatial direction). This growth factor is applied to retrieved nucleus-centered rates, resulting in total production rates. Determination of terminal production rates and correction for slit-losses using the standard Q-curve analysis incorporates mean emission intensities from diametrically opposed positions along the slit (e.g., DiSanti et al. 2001). This averages out potential asymmetries in the coma and so producing more representative overall production rates, as was demonstrated by Xie and Mumma (1996).

The water production rate was obtained from orders 26 and 27 in the KL1 setting and from order 26 in the KL2 setting, using twenty-five H$_2$O emission lines (ortho and para species), acquired during a total integration time of 60 min. (Since each setting gives a similar water production rate, we safely assume that inter-setting flux calibration is not an issue.) The total water production rate was equal to $(1.90 \pm 0.12) \times 10^{28}$ molecules s$^{-1}$ on UT 2010 July 26. In addition to water line emissions (Fig. 1), our observations revealed the presence of trace gases (Fig. 2): we detected and quantified HCN, $C_2H_6$, $CH_3OH$, and $NH_3$, and we set ($3\sigma$) upper limits for $C_2H_2$, and $H_2CO$. Production rates of all species and abundances relative to water are displayed in Table 2.

On UT 2010 September 18 we observed 10P for a total of 28 min integration time using the KL2 setting. Analysis of this dataset revealed a strong continuum but no apparent spectral emission lines. Since the rotational temperature is poorly constrained by our observations, we adopted 30 K (assuming $T_{rot}$ varies as ~ $R_h^{-1.5}$ heliocentric dependence, Biver et al. 1999) on September 18. An upper limit ($3\sigma$) for the global water production rate water was then estimated ($2.86 \times 10^{27}$ molecules s$^{-1}$, a terminal value). We assumed a representative growth factor to be 1.6, which is consistent with previous observations with NIRSPEC.

During the 1988 apparition, the reported (post-perihelion) water production rates at $R_h$ = 1.44 AU were $\sim 2 \times 10^{28}$ molecules s$^{-1}$, derived from visual magnitudes reported in IAU Circulars, and $\sim 2.9 \times 10^{28}$ molecules s$^{-1}$, based on UV observations with the International Ultraviolet Explorer (IUE) satellite (see Roettger et al. 1991, and references therein). Likewise, estimates at $R_h$ = 1.62 AU resulted in $Q(H_2O) \sim 1.5 \times 10^{28}$ molecules s$^{-1}$ (from IAU Circulars) and $(1-2) \times 10^{28}$ molecules s$^{-1}$ (IUE satellite). Our upper limit in September 2010 is lower than those values measured in 1988. Such strong decrease, however, is supported by contemporary observations in the visual[1]. Based on a sample of 37 comets, Jorda et al. (2008) derived an empirical relation between $Q(H_2O)$ and visual magnitude {log $Q(H_2O)$ = 30.675 - 0.2453 ($m_v$ – 5 log $\Delta$)}. Applied to 10P, this relation implies a decrease in $Q(H_2O)$ by a factor of at least 7 between July 26 and September 18, consistent with our measurements. Similarly, recent estimates of production rates from HST (Hubble Space Telescope) observations determined a (pre-perihelion) water production of 1.8 × $10^{27}$ molecules $s^{-1}$ on 1999 June 23 (at $R_h$ = 1.67 AU and Δ = 0.68 AU; Lamy et al., 2009), also consistent with our upper limit at $R_h$ = 1.62 AU post-perihelion.

---

[1] http://www.aerith.net/comet/catalog/0010P/2010.html. Based on these data, we adopted a visual magnitude $m_v \sim 9 \pm 0.5$ (~13.5 ± 0.5) on UT 2010 Jul 26 (on Sept 18) in our calculations.

### 3.3. OPR and nuclear spin temperature of $H_2O$

Molecules containing symmetrically located hydrogen atoms in their molecular structure, e.g., $H_2O$, $H_2CO$, are quantized by the total nuclear spin angular momentum, which depends the relative orientation of the proton spins. Basically, these species indicate the molecular spin of the hydrogen atoms, being parallel or anti-parallel, respectively. The abundance of these (ortho and para spin) species is contingent upon the temperature when the molecule was last equilibrated (Bockelée-Morvan et al. 2004), and these spin properties are not effectively modified by inelastic collisions or radiative processes in the coma (Mumma et al. 1987; Kawakita and Kobayashi 2009, and references therein). Following these molecular properties, Crovisier (1984); Mumma et al. (1987) suggested that the OPR should serve as an indicator of the thermal conditions when cometary ices agglomerated, and thus should provide perspectives on the formation region(s) within the protoplanetary disk.

To obtain the OPR in water, we used (ten) ortho-water and (four) para-water emission lines from orders 26 and 27 in the KL1 setting. Our improved quantum band model for water provided a satisfactory fit to all lines, and resulted in an OPR = 3.01 ± 0.18, corresponding to (1σ) $T_{spin}$ > 38 K. This result is consistent with complementary calculations using ortho and para $H_2O$ lines from orders 26 and 27 in the KL1 setting and order 26 in the KL2 setting (Fig. 1). Previous studies have reported spin temperatures in water (Crovisier et al. 2006) and ammonia (Kawakita et al. 2004) to be clustered near 30 K in several comets, although measurements in some comets are consistent with OPR in statistical equilibrium (OPR = 3.0, $T_{spin}$ > 50 K) (e.g., Dello Russo et al. 2008; Bonev et al. 2007). For a recent review of spin temperatures in water, ammonia, and methane, see Mumma and Charnley (2011).

### 3.4. Spatial profiles

In mid-July, the projected comet-Sun radius vector was oriented at position angle

(P.A.) 72.5° (measured CCW from North) and optical images revealed a prominent jet in CN at P.A. ~20° (Schwieterman et al. 2010). The CN jet did not vary with rotational phase or in position throughout June and July (Knight and Schleicher 2010).

On UT 2010 July 26, we oriented the slit along the jet (i.e., 20° CCW from North) in order to sample the gas activity, and to investigate spatial profiles of primary volatiles as revealed in the IR. Compared with dust, all four volatiles are much more extended in the jet direction (negative pixels along the slit, Fig. 5). $C_2H_6$, $H_2O$ and $CH_3OH$ show similar outgassing patterns, suggesting a common mechanism of release. Hydrogen cyanide is also enhanced in the jet direction, but its intensity peaks closer to the nucleus and the intensity distribution is flatter there, compared with the other gases. The Sun-Comet-Earth (i.e., phase) angle of 39.9° suggests that the strong enhancement seen in these profiles is likely associated with local insolation on the nucleus.

The maxima in the spatial intensity profiles for $C_2H_6$, $H_2O$ and $CH_3OH$ are displaced sunward from the continuum peak by about 200 km (to the left in our plot). In the anti-jet direction, these four volatiles follow a distribution similar to the dust continuum, and no strong excursions are observed. At $R_h = 1.5$ AU in 1988, Spinrad (1991) found similar displacements between the continuum and some radicals ($NH_2$, $C_2$, and [O I]) in the sunward direction. (We note that the O I emission is prompt emission following dissociation of the $H_2O$ parent, and thus the O I profile tracks that of the primary volatile.) If we take the continuum peak to indicate the nucleus position, this behavior might be explained by the existence of a distributed source mechanism, such as release of gases from icy grains that are removed from the nucleus before subliming completely.

We tentatively interpret the release of icy grains by a simple approximation, although a full analysis requires a comprehensive model of grain production. A study by Beer et al. (2006), which accounts for radiative heating, radiative cooling, and evaporative cooling to estimate the energy balance on grains in cometary comae, confirmed that sublimation rates in ice grains depend on grain size, composition, and distance from the Sun. The estimated lifetime of 1 μm (dirty) ice grains at $R_h = 1.44$ AU is $l_{grain} \sim 400$ s. Biver et al. 1999 estimate the outflow gas velocity to be $v_{gas} = 0.9\ R_h^{-0.4}$ km s$^{-1}$ (i.e. 0.78 km s$^{-1}$ at $R_h$ = 1.44 AU) at post perihelion distances. If the terminal velocity of the grain is half that of the gas, the total sublimation of these icy grains would occur by nucleocentric distances ($d_{sublimation} = v_{grain} \cdot l_{grain}$) of ~150 km. For grains of 1-cm size, the estimated lifetime is 2 × 10$^5$ s, and with a terminal velocity of 1 m s$^{-1}$ their total sublimation would occur by a nucleocentric distance of ~200 km. The sublimation of massive icy clumps, on the other hand, would impart only very small center-of-mass motion to gases sublimed from them, and the spatial profiles of such gases would peak at the same apparent position as the nucleus and be more uniformly distributed about it (such as in 103P/Hartley 2, see Mumma et al. 2011, and in C/2007 W1 (Boattini), see Villanueva et al. 2011b). A peak displacement of ~200 km from the nucleus' center, however, suggests that sublimation was somewhat delayed compared with direct (and smooth) release at the nucleus. This favors the idea of small (~micron size) particles with high outflow velocity instead of large clumps. More detailed modeling is beyond the scope of this paper.

The continuum profile decreased sharply in both directions (similar to the PSF of the flux standard star), suggestive of a point-like source. The fairly symmetric and much sharper infrared continuum could be explained by either a collimated coma (cf. Hewitt and Luu 1989), a dust-free or dust-poor coma (cf. A'Hearn et al. 1989; Roettger et al. 1990; Lamy et al. 2009), or our slit P.A. not being positioned along the dust vent (Schwieterman et al. 2010 found an offset between the CN jet and dust jet in their 1999 and 2010 optical observations). The principal contribution is likely reflected from and/or emitted by the nucleus, with some contribution possible from large refractory grains (e.g. Kelley et al., 2004).

On July 26, the estimated flux density of the continuum was (1.3 ± 0.4) × 10$^{-15}$ W m$^{-2}$ μm$^{-1}$ in the L-band (3.45 μm). Using a standard (non-spherical) thermal model and a chi-square fit, the observed flux density is well described by reflected sunlight and thermal emission from the nucleus at a nucleus temperature of 290 ± 10 K (assuming an effective radius of 4.8 km, an albedo of 0.07, and an emissivity of 0.9; further discussion of the various contributions in the 2.8–3.7 μm region and of the model used is deferred to a future publication). At $R_h$ = 1.72 AU, A'Hearn et al. (1989) estimated a nucleus temperature of 270 K. These temperatures exceed those of a blackbody at these heliocentric distances, suggesting some contribution from dust. We note, however, that surface temperatures larger than 300 K have been estimated on several comets (at similar $R_h$) by in-situ measurements (e.g., Soderblom et al. 2004; Groussin et al. 2007). On 1989 October 4.67 ($R_h$ = 1.39 AU, post perihelion), Tokunaga et al. 1989 measured a nucleus flux of 0.82 × $10^{-15}$ W $m^{-2}$ $μm^{-1}$ and dust flux of 5.12 × $10^{-15}$ W $m^{-2}$ $μm^{-1}$ in the K-band (2.19 μm). Based on these estimates, our measured flux seems to be consistent with a contribution mostly from the nucleus.

The spatial profiles for primary volatiles are strongly enhanced in the direction of a jet observed at optical wavelengths in 2010 mid-July (M. Knight 2011, private communication). A comparison of the two data sets will help to address the possible role of HCN as progenitor for the CN radical (e.g., Paganini et al. 2010).

## 4. Discussion

Jupiter Family comets (JFCs) have experienced multiple apparitions since the first direct detections of primary volatiles (in 1985) and the advent of suitably sensitive astronomical facilities (1990–2000). Such warming is a common occurrence for many comets in short period orbits, and for them it is not yet certain whether the observed abundances of native species reflect the formation heritage of cometary ices or instead are a consequence of thermo-chemical evolution during multiple perihelion passages. (Only a dozen JFCs have been characterized even once in primary volatiles at infrared wavelengths, and even fewer on multiple occasions.) Several studies, however, strongly favor the idea that their observed volatile compositions might indeed preserve intrinsic characteristics from the epoch of formation:

- Production *rates* can vary greatly on time scales from hours (generally associated with nucleus rotation) to months (associated with orbital motion), but the relative abundance *ratios* and physical properties in released volatiles may remain unchanged. The production *rates* of primary volatiles in JFC 103P/Hartley-2 varied by a factor of two with nucleus rotation, but their abundance *ratios* remained constant to a small degree of uncertainty both day-by-day and over a span of three months during the 2010 apparition (Mumma et al. 2011; Dello Russo et al. 2011). Again, in 1P/Halley (a nearly isotropic comet of OC origin), the relaxed nuclear spin temperature was unchanged over a span of three months (December 1985 – March 1986) even though the comet experienced massive surface wasting during that time and was characterized by multiple active regions (Mumma et al. 1993; Bockelée-Morvan et al. 2004).

- Fragments B and C of the split JFC 73P/SW3 showed quite different production *rates*, but essentially identical chemical *properties*, including severe depletion in organic volatiles (Villanueva et al. 2006; Dello Russo et al. 2007; Kobayashi et al. 2007; Paganini et al. 2010). Since fragment B was undergoing complete disintegration, the comparison demonstrated that the two fragments shared a common chemical composition and history, independent of any earlier surface processing.

- The chemistry of JFC 21P/Giacobini-Zinner remained unchanged on successive apparitions (1998 and 2005), showing that the composition of its active layers did not vary with depth in the nucleus despite significant surface wasting (Weaver et al. 1999; Mumma et al. 2000; DiSanti et al. 2012, in prep.).

- In particular, outburst events might offer a genuine opportunity to examine relatively pristine constituents from the bulk material released from the inner nucleus, where such material had little (if any) chance of experiencing chemical processing by thermal and radiative effects during the comet's lifetime. The enriched organics and very low spin temperature found for $H_2O$ in the outbursting OC comet C/2001 A2 (LINEAR) may be especially significant. By contrast, the disintegrating JFC 73P/Schwassman-Wachmann 3 showed depleted organics and nuclear spin ratios consistent with statistical equilibrium, for $H_2O$ and $NH_2$ (i.e., $NH_3$).

The diverse composition of organics in comets has led IR observers to suggest a chemical taxonomy based on organic content and composition relative to water vapor, including organics-enriched, -normal and -depleted populations (Mumma et al. 2003). Infrared observations to date have suggested that most JFCs are relatively depleted in organic volatiles with respect to water vapor, although this is based on a very small number of comets. In comet 10P, the abundance ratios (relative to water) for $CH_3OH$ and $NH_3$ are consistent with organics-normal comets, $C_2H_6$ and HCN are significantly depleted, and $C_2H_2$ and $H_2CO$ are severely depleted. Other exceptions to the trend of JFCs being "organics-depleted" have been found—material excavated from the nucleus of 9P/Tempel 1 by Deep Impact was found to be normal in the ratio $C_2H_6/H_2O$ (Mumma et al. 2005, DiSanti et al. 2007); material ejected from 17P/Holmes during a massive outburst in 2007 was enriched in organics (Dello Russo et al. 2007); and 103P/Hartley-2 had nearly-normal organic abundances (A'Hearn et al. 2011; Mumma et al. 2011; Dello Russo et al. 2011). This diversity strengthens the view that comets might not have a common heritage, and thus it supports current dynamical evolution models that suggest accretion occurred in different regions of the proto-solar nebula (see Duncan 2008, and references therein).

A taxonomy based on organic volatiles in the radio demonstrated no clear trends among JFCs and Oort Cloud comets (Crovisier et al. 2009). However, compositional comparisons based on radio observations could be biased owing to technical considerations (see the review by Mumma and Charnley 2011). Figure 6 (and Table 3) displays a comparison of mixing ratios for primary volatiles in JFCs from IR observations. Based on the current sample of eight JFCs, we observe a similar trend in composition (though methanol displays strong deviations), suggesting a common place of formation (with a few exceptions as mentioned above). Needless to say, a larger sample will reinforce current statistics and shed further light on this discussion. We note that, although the current database of JFCs at infrared wavelengths totals fewer than a dozen comets, the recent advent of new state-of-the-art IR facilities has permitted its dramatic expansion since 1999.

When connected with the chemical composition of primary volatiles, abundance ratios of nuclear spin species (i.e., ortho-para ratios) and spin temperatures may aid in the understanding of cometary origins. Questions have been raised whether or not values for these result from alteration in the cometary nucleus (e.g. Kawakita et al. 2004), or might result from the interaction of gases with dust grains. Such questions are as yet unresolved, largely because laboratory measurements are difficult and lacking (Crovisier 2007; Bonev et al. 2007; Kawakita and Kobayashi 2009, and references therein, however see Mumma et al. 1987, and Mumma and Charnley 2011). External factors are believed to modify these physical properties in the meter-thick outer layer of the cometary nucleus, for example solar irradiation of JFCs during multiple perihelion passages and cosmic-ray irradiation of Oort Cloud comets (Mumma et al. 1988). If these catalysts were indeed negligible, the OPR analysis should provide hints to the initial characteristics and formation conditions of cometary volatiles. The relatively low abundance of organics and an OPR consistent with statistical equilibrium in comet 10P suggest that accretion of its cometary ices occurred in the proto-planetary disk, closer to the Sun, where some degree of thermal processing occurred.

Among the organic compounds measured, we find a relatively high abundance of ethane and methanol with respect to acetylene and formaldehyde, respectively. The ratio $C_2H_6/(C_2H_2 + C_2H_6) > 0.86$ observed in comet 10P evokes the idea of formation via hydrogenation processes (Mumma et al. 1996). Similarly, an indirect evaluation of CO-ice hydrogenation can be tested through a high abundance of $CH_3OH$ with respect to $H_2CO$, following the $CO \rightarrow H_2CO \rightarrow CH_3OH$ chemical reaction (Watanabe et al. 2003; Fuchs et al. 2009). (Note that with single aperture measurements, we quantify $H_2CO$ released as a monomer but we cannot test the abundance of polymeric formaldehyde. Thus, our ratio $CH_3OH/(H_2CO + CH_3OH)$ provides only an upper bound (< 95%) to the hydrogenation efficiency of formaldehyde in pre-cometary ice.) Hydrogenation of simple hydrocarbons (e.g. $C_2H_2$, $C_2H_4$) is a likely process at low temperatures (Hiraoka et al. 2000). Hence, the legacy of pre-cometary ices has been linked to interstellar chemistry because of the intrinsic low temperature in the ISM, wherein adsorbed H-atoms have higher sticking probability and less chance for recombination reactions ($H + H \rightarrow H_2$) on the grain.

Hydrogenation, which takes place at low temperatures (e.g., < 20 K), is not compatible with spin temperatures above 38 K (from our OPR estimate), because the latter indicates formation in a warm environment. If these two measurements reveal the "true" (and undisputed) history of cometary nuclei during formation, then either they occurred at different epochs, or alternatively the measured cometary ethane and methanol might have been produced by sources other than hydrogenation. Indeed, ubiquitous presence of methanol and methane has been found in cold regions of Young Stellar Objects (Boogert et al. 2008; Boogert et al. 2011, and references therein). These findings suggest other possible sources for $C_2H_6$ and $CH_3OH$ in pre-cometary ices, where irradiated methane ice might serve as a source of ethane by the exothermic reaction of two methyl radicals (Hudson & Moore 1997; Bennett et al. 2006).

5.   **Conclusions**

We observed the Jupiter Family Comet 10P/Tempel 2 on UT 2010 July 26 and September 18, using high-dispersion infrared spectroscopy. On July 26, we detected water (production rate $(1.90 \pm 0.12) \times 10^{28}$ molecules $s^{-1}$; rotational temperature $35 \pm 3$ K), and multiple trace gases. The rotational temperature for HCN ($36^{+13}_{-9}$ K) was consistent with that found for $H_2O$. We did not detect molecular emission features in September, but we did obtain a sensitive upper limit for water production ($2.86 \times 10^{27}$ molecules $s^{-1}$, assuming $T_{rot} = 30$ K).

Spatial profiles in comet 10P revealed significant release of major volatiles in a direction (P.A. 20º) coincident with a jet-like feature observed by optical observers in 2010 mid-July. The particular off-centered profiles displayed by these volatiles suggested the possible existence of a distributed source, such as the release of icy grains that subsequently sublimed in the coma.
Provided its composition revealed pristine conditions during the perihelion passage in 2010, our measured volatile composition in comet 10P showed evidence of formation in a "warm" environment (a relative low organic composition and OPR consistent with statistical equilibrium). However, hydrogenation of certain volatiles, which suggests processing in a "colder" region, could have occurred at an earlier stage. This issue is a major focus of astrochemical models predicting the processing history of volatiles from their (suggested) varied formation environments ranging from proto-solar clouds to the mid-plane of subsequently evolved proto-planetary disks where comets later formed.

**Acknowledgments.**

Keck telescope time was granted by the University of Hawaii and NOAO (through the Telescope System Instrumentation Program funded by NSF). L.P. thanks two anonymous referees and Michael S. Kelley for their useful comments, and gratefully acknowledges support from the NASA Postdoctoral Program. We also acknowledge support by the NSF Astronomy and Astrophysics Research Grants Program (AST-0807939, PI Bonev), by the NASA Astrobiology Institute (RTOP 344-53-51, PI Mumma; NNA09DA77A, PI Meech), and by NASA's Planetary Astronomy (RTOP 344-32-98, PI DiSanti; 08-PAST08-0034, PI Villanueva), and Planetary Atmospheres Program (RTOP 344-33-55, PI DiSanti; 08-PATM08-0031, PI Villanueva). We thank the W.M. Keck Observatory staff for their support, especially Scott E. Dahm for his friendly advice and assistance. The authors acknowledge the very significant cultural role and reverence that the summit of Mauna Kea has always had within the indigenous Hawaiian community.

Table 1. Log of Observations

| Date (UT 2010) | $R_h$[a] (AU) | $\Delta$[b] (AU) | $\Delta_{dot}$[c] (km s$^{-1}$) | P.A.[d] (°) | $\alpha$[e] (°) | Setting[f] | $I_{time}$[g] (min) |
|---|---|---|---|---|---|---|---|
| July 26 | 1.441 | 0.687 | -4.10 | 252.5 | 39.9 | KL1/KL2 | 32/28 |
| Sept 18 | 1.623 | 0.684 | 5.34 | 297.9 | 19.5 | KL2 | 28 |

a. Heliocentric distance
b. Geocentric distance
c. Relative geocentric velocity of the comet
d. Solar Position angle, measured counter clockwise from North.
e. Solar Phase angle.
f. NIRSPEC instrument configuration.
g. Total on-source integration time.

Table 2. Molecular Parameters for Primary Volatiles in 10P/Tempel 2 on UT 2010 July 26*

| Molecule | Time (UT) | Setting/ Order [a] | $T_{rot}$ (K) | Lines | Flux ($10^{-20}$ W/m$^2$) | g-factor (phot s$^{-1}$ molec$^{-1}$) | NC Q [b] ($10^{24}$ s$^{-1}$) | GF [c] | Global Q [d] ($10^{24}$ s$^{-1}$) | Abundance % |
|---|---|---|---|---|---|---|---|---|---|---|
| $H_2O$ | 13:19–14:15 | KL1/27 | 35 ± 3 | 8 | 77.8 ± 1.1 | 5.26E-06 | 11337 ± 171 [e] | 1.7 | 19272 ± 1170 | 100 |
| | " | KL1/26 | " | 7 | 11.9 ± 0.3 | 8.41E-08 | 11173 ± 397 | 1.7 | 18999 ± 1896 | 100 |
| | 14:27–15:23 | KL2/26 | " | 10 | 16.0 ± 0.4 | 1.08E-06 | 11528 ± 284 | 1.6 | 18444 ± 1239 | 100 |
| $C_2H_6$ | 13:19–14:15 | KL1/23 | (35) | 11 | 12.6 ± 0.5 | 2.59E-04 | 43.9 ± 3.5 | 1.7 | 74.7 ± 7.0 | 0.39 ±0.04 |
| $CH_3OH$ | " | KL1/22 | " | 10 | 7.2 ± 0.6 | 3.80E-05 | 179.2 ± 21.3 | " | 304.7 ± 40.4 | 1.58 ± 0.23 |
| HCN | 14:27–15:23 | KL2/25 | 36 (+13, -9) [f] | 7 | 3.93 ± 0.34 | 2.09E-04 | 15.2 ± 1.8 | 1.6 | 24.4 ± 3.2 | 0.13 ± 0.02 |
| $NH_3$ | " | " | (35) | 2 | 0.74 ± 0.16 | 6.64E-06 | 95.5 ± 20.9 | " | 153 ± 34 | 0.83 ± 0.20 [g] |
| $C_2H_2$ | " | " | (35) | 7 | < 0.007 | 1.22E-04 | < 7.67 | " | < 12.3 | < 0.07 |
| $H_2CO$ | " | KL2/21 | (35) | 18 | < 4.72 | 2.11E-04 | < 12.7 | " | < 20.3 | < 0.11 |

* Mixing ratios are expressed relative to $H_2O$. Uncertainties represent 1σ, and upper limits represent 3σ. Please see Table 4 (in electronic form) for detailed version of this table.

a. NIRSPEC instrument setting.
b. Nucleus-centered (NC) production rate.
c. Growth factor.
d. Global production rate, after applying a measured growth factor (and its error of ± 0.1) to the NC production rate.
e. The reported error in production rate includes the line-by-line scatter in measured column densities, along with photon noise, systematic uncertainty in the removal of the cometary continuum, and (minor) uncertainty in rotational temperature.
f. We tabulate the retrieved $T_{rot}$ and confidence limits; however, all measured temperatures are consistent with 35 K. For HCN and $C_2H_6$, the derived Q's are only weakly sensitive to $T_{rot}$, so we adopted 35 K for all species when calculating the NC production rates.
g. Fink (2009) estimated $NH_2/H_2O$ equal to 0.19% on 1988 Oct 9 ($R_h$ = 1.41 AU, Δ = 1.07 AU). If we apply fluorescence efficiencies from Kawakita and Watanabe (2002) to Fink's estimates, the $NH_2$ abundance increases to ~0.4%, which is a factor of 2 lower than that required to fulfill $NH_3$ as the main source for $NH_2$ (photodissociation of ammonia leads to direct formation of $NH_2$, with 95% efficiency).

Table 3. Mixing Ratios for Primary Volatiles in JFCs, based on IR Observations*

| | | Year | $C_2H_6$ | $C_2H_2$ | $CH_3OH$ | §$CH_3OH$ | $H_2CO$ | HCN | $NH_3$ |
|---|---|---|---|---|---|---|---|---|---|
| 21P/G-Z | | 1998 | 0.22±0.13[a] | < 0.42[b] | 0.9–1.4[b] | 0.9–1.4 | < 0.8[b] | < 0.27[b] | – |
| 9P/Tempel 1 | Pre-impact | 2005 | 0.23±0.04[c] | – | 1.3±0.2[d] | 1.4±0.2 | – | 0.18±0.06[d] | – |
| " | †Post-impact | 2005 | 0.40±0.04[c] | 0.13±0.04[c] | 0.99±0.17[c] | 1.06±0.2 | – | 0.22±0.03[c] | – |
| " | Ejecta | | 0.55±0.09[c] | – | – | – | – | 0.24±0.07[c] | – |
| 73P/SW3-B | April | 2006 | < 0.3[e] | – | – | – | – | <0.2[e] | – |
| " | May | 2006 | 0.14±0.01[f] | < 0.06[f] | 0.18±0.03[f] | 0.25±0.04 | 0.15±0.05[f] | 0.29±0.02[f] | < 0.34[f] |
| 73P/SW3-C | April | 2006 | 0.15±0.04[e] | 0.23±0.06[e] | < 0.51[e] | < 0.38 | < 0.51[e] | 0.21±0.04[e] | – |
| " | May | 2006 | 0.11±0.01[f] | < 0.033[f] | 0.15±0.03[f] | 0.21±0.04 | 0.95±0.02[f] | 0.21±0.01[f] | < 0.29[f] |
| 17P/Holmes | | 2007 | 1.78±0.26[g] | 0.34±0.01[g] | 2.25±0.43[g] | 3.15±0.6 | 0.54±0.08[g] | – | – |
| 6P/d'Arrest | | 2008 | 0.26±0.06[h] | < 0.05[h] | 1.42±0.3[h] | 1.99±0.42 | 0.36±0.09[h] | 0.03±0.01[h] | 0.52±0.15[h] |
| 103P/Hartley 2 | | 2010 | 0.75±0.03[i] | 0.11±0.02[i] | 2.84±0.50[i] | 2.13±0.38 | 0.33±0.10[i] | 0.22±0.01[i] | 0.56±0.15[i] |
| " | | 2010 | 0.65±0.09[j] | 0.14±0.02[j] | 1.25±0.14[j] | 1.75±0.2 | 0.11±0.02[j] | 0.27±0.03[j] | 0.84±0.16[j] |
| 10P/Tempel 2 (this work) | July | 2010 | 0.39±0.04 | < 0.07 | 1.58±0.23 | 1.58±0.23 | < 0.11 | 0.13±0.02 | 0.83±0.20 |

* Mixing ratios are expressed relative to $H_2O$. Uncertainties represent 1σ, and upper limits represent 3σ.

† The Post-impact values represent the sum of quiescent release and ejecta (see Mumma et al. 2005).

§ We use a $CH_3OH$ $v_3$ Q-branch g-factor of $1.5 \times 10^{-5}$ $s^{-1}$ from Villanueva et al. (2011c) to normalize previous studies using different fluorescence models.

a. Mumma et al. 2000.

b. Weaver et al. 1999.

c. DiSanti and Mumma 2008.

d. Mumma et al. 2005.

e. Villanueva et al. 2006.

f. Dello Russo et al. 2007. Values are consistent with independent measurements in 73P-B by Kobayashi et al. (2007).

g. Dello Russo et al. 2008.

h. Dello Russo et al. 2009.

i. Mumma et al. 2011.

j. Dello Russo et al. 2011.

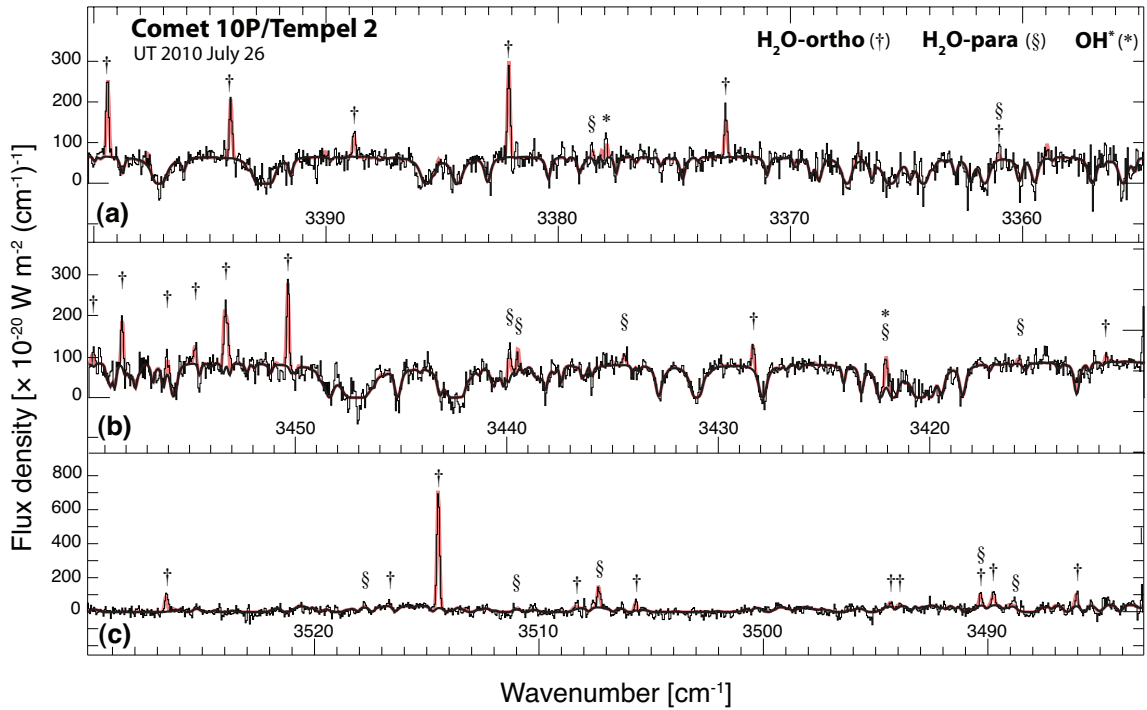

Figure 1. Water emission lines detected in comet 10P/Tempel 2 on UT 2010 July 26. OH* prompt emission lines are also identified within these spectral bands. Simultaneous detection of water lines facilitated the retrieval of the rotational temperature, production rate, and nuclear spin parameters (spin temperature). Emission lines detected in different grating settings: (a) KL1 setting, order 26, (b) KL2 setting, order 26, and (c) KL1 setting, order 27. Here, and in Fig. 2, the continuous black line underlying each spectrum represents a model of the atmospheric transmittance function (convolved to the spectral resolving power of our observations, then scaled to the cometary continuum), and the continuous red line depicts the modeled spectra.

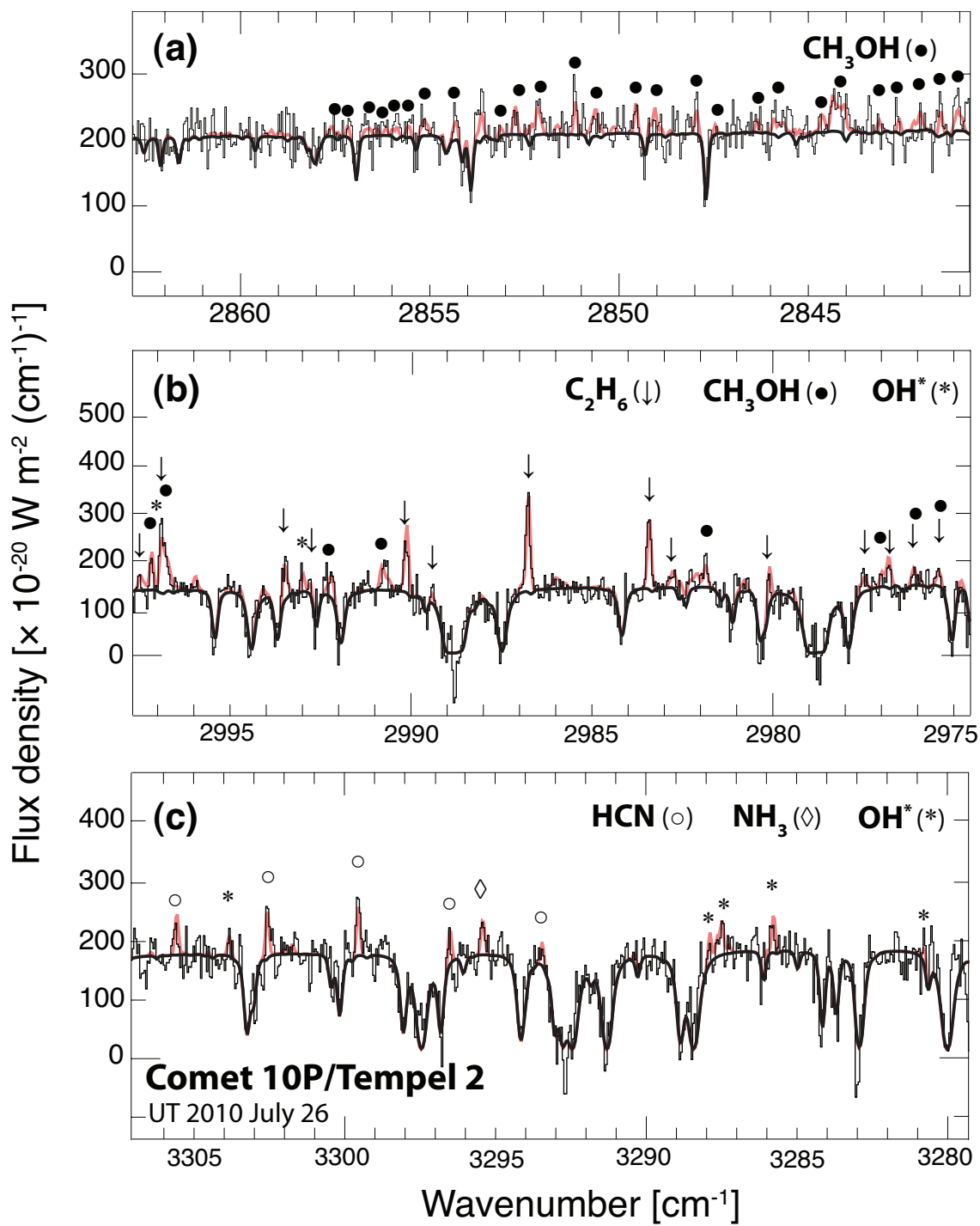

Figure 2. Astronomical observations of comet 10P unveiled the presence of minor volatiles on UT July 26. Panel (a) displays the methanol $\nu_3$ band in order 22 (KL1). (b) The 2975–3000 cm$^{-1}$ band (KL1 - order 23) covers ethane $\nu_7$ (P, R, and Q-lines); some of these lines are blended with $CH_3OH$ and OH*. (c) Order 25 (KL2) shows HCN (P-branch lines) along with $NH_3$ and OH*. Our analysis of HCN emission line intensities resulted in a $T_{rot}$ of $36\,^{+13}_{-9}$ K (see Fig. 3).

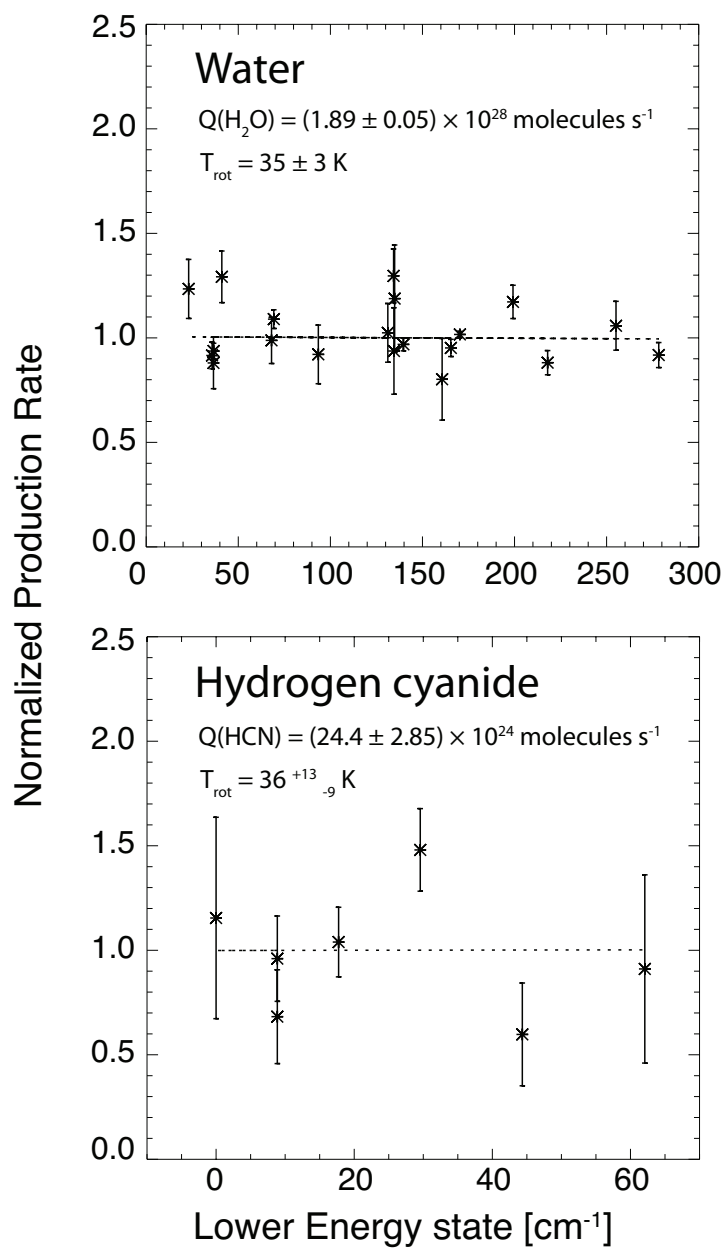

Figure 3. Excitation analysis for $H_2O$ (upper) and HCN (lower) on 2010 July 26. The line fluxes are obtained from nucleus-centered extracts of 3 by 9 pixels (equivalent to 0.43″ × 1.78″, or 266 × 800 km) and are corrected for the monochromatic terrestrial transmittance at each Doppler-shifted line frequency.

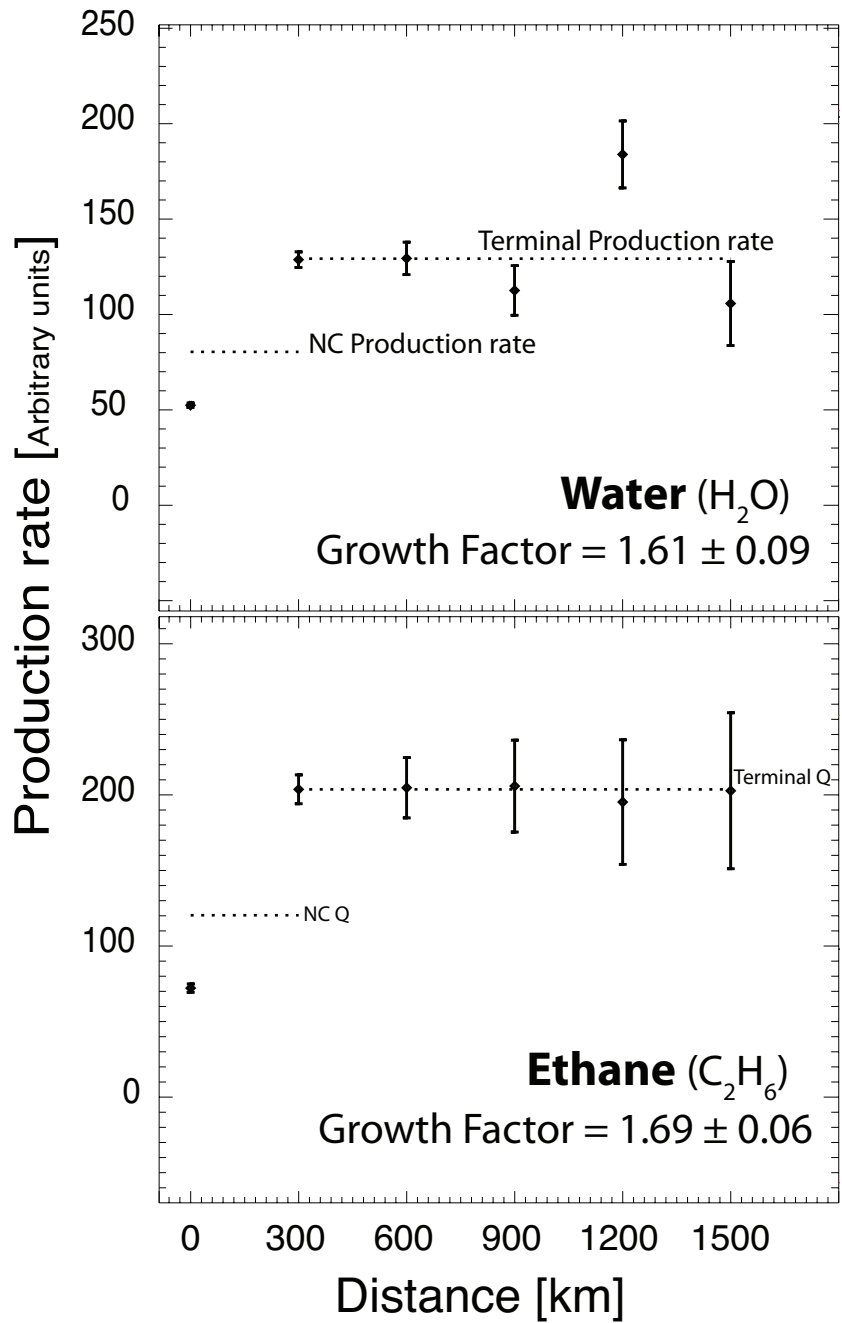

Figure 4. Growth factors for water (upper) and ethane (lower). The growth factor describes the ratio of terminal to nucleus-centered production rates, where terminal refers to off-nucleus extracts along the slit spatial direction (averaged at corresponding distances to either side of the nucleus).

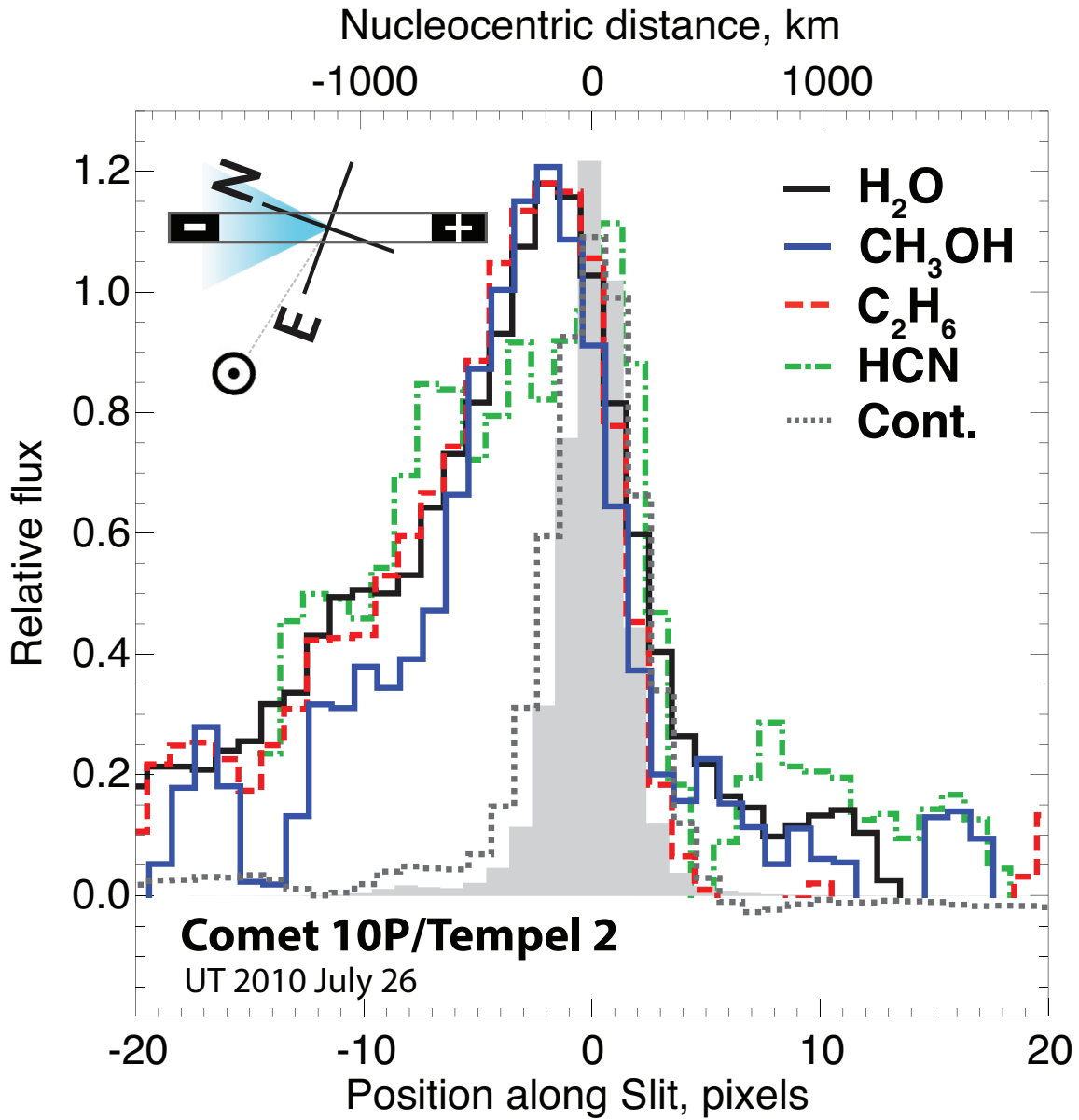

Figure 5. Spatial profiles of primary volatiles and continuum in comet 10P. These profiles reveal significant outflow of major volatiles at a P.A. = 20°, which was coincident with a jet-like feature observed by optical observers in 2010 mid-July. We aligned the slit along the jet, which is placed horizontally in this plot. The ends of the slit (+, -) indicate the direction of pixels displayed on the abscissa and the grey shaded region is a representation of the stellar PSF. See Section 3.4.

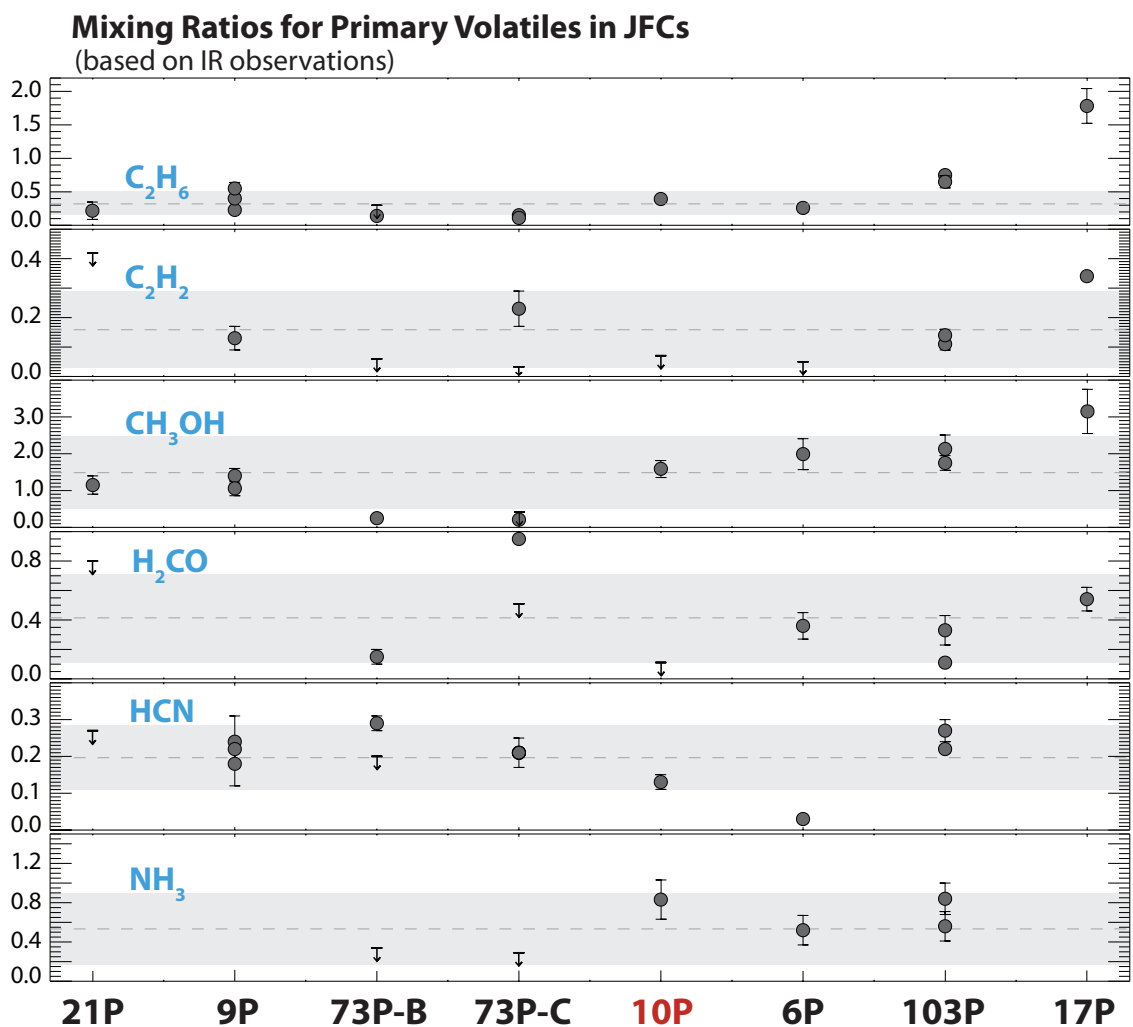

Figure 6. Mixing ratios are based on current IR observations of eight JFCs. The dashed line shows the standard mean of the sample and the shaded grey region represents its 95% confidence interval (which is not a prediction interval). In order to derive these values we have: a) assumed a Gaussian distribution of abundances for each species, b) used the mean value if more than one measurement exists for a specific molecule in a comet, and c) considered upper limits (as values) in our sample. For $CH_3OH$, we provide normalized values to account for different g-factors used by previous studies (see note '§' in Table 3). These observations are described in Table 3 and discussed in Section 4.